\documentclass[doublecol]{epl2} 
\pdfoutput=1
\usepackage{bm}
\usepackage{amsmath}
\usepackage{graphicx}
\usepackage{amssymb}
\usepackage{esint}
\makeatletter
\newcommand{\acosh}{{\rm arccosh\,}}

\title{Coulomb Drag and High Resistivity Behavior in Double Layer Graphene}

\author{N. M. R. Peres\inst{1}
\and J. M. B. Lopes dos Santos\inst{2}
\and A. H. Castro Neto\inst{3,4}}

\institute{
\inst{1} Department of Physics and Center of Physics, University of
Minho, P-4710-057, Braga, Portugal, EU

\inst{2} CFP and Departamento de F\'{\i}sica, Faculdade de Ci\^{e}ncias
Universidade do Porto, P-4169-007 Porto, Portugal, EU

\inst{3} Graphene Research Centre and Physics Department,
National University of Singapore, 2 Science Drive 3, Singapore 117541

\inst{4} Department of Physics, Boston University, 590 Commonwealth Avenue,
Boston, MA 02215, USA
}


\abstract{We show that Coulomb drag in ultra-clean graphene double layers can be used 
for controlling the on/off ratio for current flow by tunning the external gate voltage. 
Hence, although graphene remains semi-metallic, the double layer graphene system can be 
tuned from conductive to a highly resistive state. We show that our results explain 
previous data of Coulomb drag in double layer graphene samples in disordered SiO$_2$ substrates.}

\begin{document}
\maketitle

\section{Introduction}

\begin{figure}[ht]
 \begin{center}
   \includegraphics*[width=6cm]{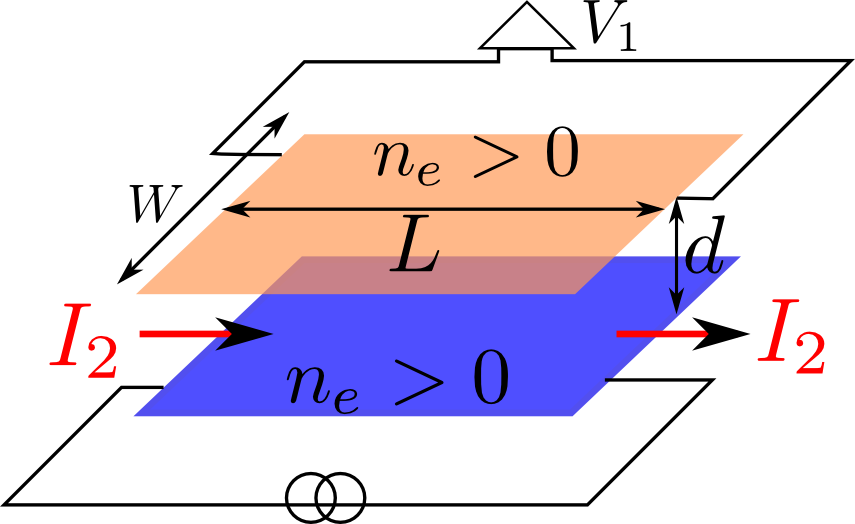}
 \end{center}
\caption{Generic representation of a Coulomb drag experiment.}
\label{fig_drag_exp}
\end{figure}

Coulomb drag, as represented in Fig. \ref{fig_drag_exp}, is the phenomenon 
where a voltage $V_2$ applied to a two-dimensional 
(2D) conducting layer (called the active layer) generates 
both 
a current $I_2$ on that plane and a voltage $V_1$ in another layer, parallel to the first,
and located at a distance $d$ \cite{Rojo}. 
This effect occurs because electrons in the active layer, 
under the presence of an electric field $E_2$, ``drag'' the electrons in the other 
layer through their mutual Coulomb 
interaction. The current densities in each layer, $j_1$ and $j_2$, are related to 
electric fields in each layer through 
the conductivity tensor, according to the general relation: 
\begin{equation}
\left[
\begin{array}{c}
 j_1\\
j_2
\end{array}
\right] =
\left(
\begin{array}{cc}
 \sigma_{11}&\sigma_{D}\\
\sigma_{D} &\sigma_{22}
\end{array}
\right)
\left[
\begin{array}{c}
 E_1\\
E_2
\end{array}
\right]\equiv\bar\sigma
\left[
\begin{array}{c}
 E_1\\
E_2
\end{array}
\right]
\,,
\end{equation}
where $\sigma_{ii}$ is the conductivity of each isolated layer ($d \to \infty$) and 
$\sigma_D$ is the so-called trans-conductivity. Notice that in a drag experiment  no current flows
in the non-active sheet, that is, $j_1=0$, and hence we can express the electric fields $E_1$ and $E_2$ in terms of the
current $j_2$ in the active plane alone. This 
allow us to define the quantities of experimental interest, namely
the drag resistivity $\rho_D$, given by
\begin{equation}
 \rho_D =\frac{WV_1}{I_2L} = \frac{E_1}{j_2} = -\frac{\sigma_{D}}{{\rm det \bar\sigma}}\,,
\end{equation}
and the longitudinal resistivity $\rho_L$, reading
\begin{equation}
\rho_L = \frac{E_2}{j_2} = \frac{\sigma_{11}}{{\rm det \bar\sigma}} \, ,
\end{equation}
where ${\rm det \bar\sigma} = \sigma_{11} \sigma_{22} - \sigma_D^2$ is the determinant of conductivity tensor $\bar\sigma$ 
($W$ is the width and $L$ is the length of the device). 

Using the above equations we can eliminate the trans-conductivity in terms of $\rho_D$ 
and rewrite the relationship between the drag and the longitudinal resistivities as:
\begin{equation}
 \rho_L=\frac{2 \rho_D^2/\rho_1}{\sqrt{1 + 4 \rho_D^2/(\rho_{1}\rho_{2})}-1}\,,
\label{eq_rho0}
\end{equation}
where $\rho_i = 1/\sigma_{ii}$. Eq. (\ref{eq_rho0}) has two extreme limits:
\begin{equation}
\rho_L\approx
\left\{
\begin{array}{c}
\rho_{2} + \rho_D^2/\rho_1 \,,\hspace{1cm} \rho_D \ll \sqrt{\rho_{1}\rho_{2}}/2 \,, \\
\sqrt{\rho_{2}/\rho_{1}}\rho_D\,,\hspace{1.cm}
\rho_D \gg \sqrt{\rho_1 \rho_2}/2\,.
\end{array}
\right. 
\label{eq_rho0_2_limits}
\end{equation}
Hence, in the limit where the drag resistivity is small compared with the resistivity of the two layers
($\rho_D \ll \sqrt{\rho_{1}\rho_{2}}/2$) the longitudinal resistivity is dominated by the resistivity of the active layer. 
This is the usual case of the 2D electron gases (2DEG), found in semiconducting heterostructures 
\cite{Solomon89,Gramila1991,Gramila1994}, where the drag resistance is a small effect. 

In what follows, we will argue that for the case of an ultra-clean
graphene double layer the opposite regime can be reached, namely, one can have $\rho_D \gg \sqrt{\rho_{1}\rho_{2}}/2$
with $\rho_1 \ll \rho_2$ so that the longitudinal resistivity, $\rho_L$, is proportional to the drag resistivity,
but  enhanced by a factor proportional to $\sqrt{\rho_2/\rho_1} \gg 1$, so that $\rho_L \gg \rho_D \gg  
\sqrt{\rho_{1}\rho_{2}}/2$. This regime can be reached in ultra-clean graphene by tuning the bottom and top gates in
the device (see Fig. \ref{fig_drag_exp}) such that the Fermi energy of the non-active layer, $\epsilon_{F1} = v_F k_{F1}$ 
(where $v_F \approx 10^6$ m/s is the Fermi velocity), is large, but the Fermi energy of the 
active layer, $\epsilon_{F2} = v_F k_{F2}$, is small and close to the neutrality point 
(that is, one requires that $k_{F1} \gg k_{F2}$). This is only possible in graphene because the resistivity in each
individual graphene layer is a monotonically decreasing function of the density, $|n_e|$, in each plane 
($k_{Fi} = \sqrt{\pi |n_e|}$) \cite{nov04}. In ultra-clean graphene, the resistivity is a very sharp 
function of the density and therefore the
longitudinal resistivity can be enhanced by orders of magnitude by the drag effect. 
In fact, we show that the drag resistance
diverges at low densities as $\rho_D\sim n_e^{-0.8}$ ($n_e \to 0$) and therefore the longitudinal resistance can become arbitrarily large.
We stress that the behavior for $\rho_D$, as computed using our model, is better described
by the relation $\rho_D\sim n_e^{-\alpha(n_e)}$, where the exponent $\alpha(n_e)$ is a function of $n_e$ and shows a crossover
from $\alpha\simeq2$ to $\alpha\simeq1$ as $n_e$ decreases.

The mechanism described here can produce a huge enhancement of the on-and-off ratio for current flow in graphene, 
without the need for the opening of a gap in the spectrum,  solving the famous bottleneck for the use of graphene devices  in high end electronic applications. 

While we wait for the drag data in ultra-clean samples \cite{geim}, 
we check our model against  
the existing data for drag resistivity measurements in graphene on SiO$_2$ \cite{Tutuc2010}. As it is well known, SiO$_2$ is
a dirty substrate, and at low densities electron-hole puddles are formed \cite{yacoby,Cromieinhomogeneous}, 
greatly limiting the mobility
in these devices. We show that our model applies to this conventional case describing the data extremely well. 

We notice that the theoretical literature on Coulomb drag in 
graphene is scarce \cite{Tse07,Narozhny07}, and 
it was shown \cite{Tutuc2010} that the theoretical approaches 
used so far 
\cite{Tse07,Narozhny07} are unable to describe the experimental data. 
Under a number of simplifying assumptions, it was shown \cite{Narozhny07} that electrons 
in graphene, when described by the massless Dirac equation, should have zero drag-resistivity.
According to that analysis \cite{Narozhny07}, trigonal warping corrections would be necessary
to explain a finite drag resistivity in graphene. We show here that such is not the case
when the momentum dependence of the scattering time is taken into account and 
dynamic effects in Coulomb screening  are correctly included.

\section{Theory of Coulomb drag and experimental results}
\label{sec_main}

The drag resistivity can be obtained from the solution of Boltzmann's kinetic equation \cite{Solomon90,Solomon91,Jauho93,Flensberg94,Flensberg95}. In this approach it is assumed that the main scattering mechanism within a graphene layer is electron-impurity 
scattering \cite{RMP10,reports} and that the electronic density is outside the range where one finds electron-hole puddles\cite{yacoby,Cromieinhomogeneous}. In dirty substrates such as SiO$_2$ this can happen at densities of 10$^{12}$ cm$^{-2}$, however, in cleaner
substrates such as Boron-Nitride \cite{BNKim} this only happens at extremely low densities of the order of 10$^{10}$ cm$^{-2}$ 
\cite{Sarma11}. Without loss of generality we also assume that the graphene layers are electron doped and, as explained above,  the Fermi energy in the two graphene layers is such that $\epsilon_{F2}\leq\epsilon_{F1}$. We also make use of the full
dynamical screening between the layers, which takes into account intra- as well as inter-layer interactions. However, in calculating the resistivity we only take into account  intra-band interactions between electrons belonging to each of the sheets
(the validity of this assumption is achieved by keeping the electronic density large enough \cite{Narozhny07}). Finally, one  key point of our approach is  taking into account the full momentum dependence of the electronic scattering time $\tau_{\bm k}$, originated from 
electron-impurity scattering. It is well known that $\tau_{\bm k}$
is roughly proportional to the square root of the electronic density, that is, we have  $\tau_{\bm k}=\tau_0k$, where
$\tau_0$ is a constant  computed elsewhere \cite{RMP10},  and which
drops out at the end of the calculation.
This assumption is in agreement with the experimental data \cite{nov04}
and is essential for the accurate description of Coulomb drag in graphene.

\begin{figure}[ht]
 \begin{center}
\includegraphics*[width=8cm]{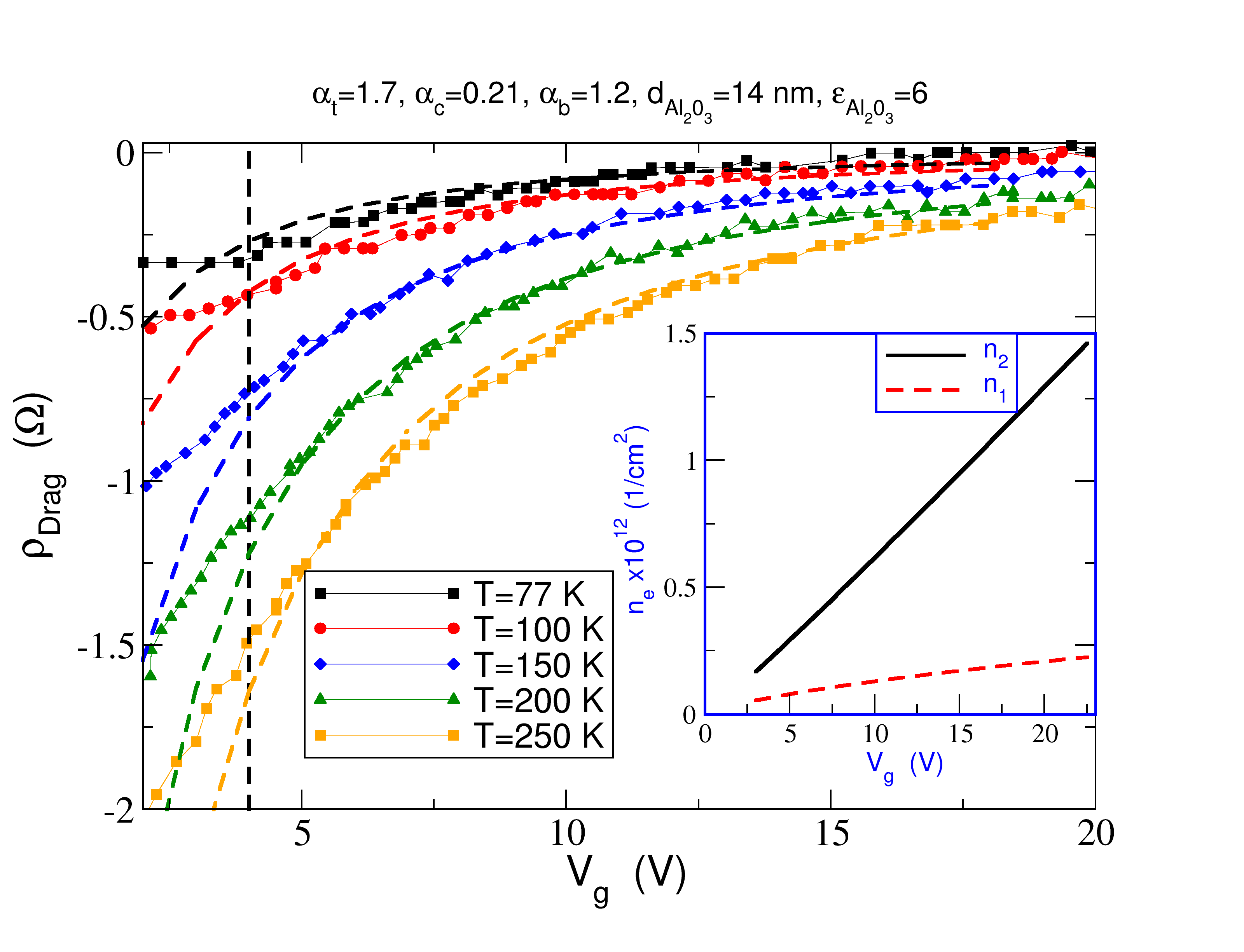}
 \end{center}
\caption{Calculation of the drag resistivity as function of
the gate-voltage and comparison with the 
experimental measurements \cite{Tutuc2010},
 using $\alpha_c=0.21$, $d_b=280$ nm, $\epsilon_b=3.8$, $d_t=14$ nm, and $\epsilon_t=6$.
Note that except for $\alpha_c$, the parameters $d_b$, $d_t$, $\epsilon_b$, and $\epsilon_t$
are a characteristic of the apparatus used in the experiment, and therefore fixed numbers. 
The symbols are the experimental data and 
the dashed lines
are the calculated values of $\rho_D$. 
The inset represents the electronic density, $n_1$ and $n_2$, in the 
two sheets, as obtained from the solution of Eqs. (\ref{eq_electro_1}) and (\ref{eq_electro_2}).
 We note that $n_1$ and $n_2$ are not independent of each 
other.}
\label{fig_drag_results}
\end{figure}

Within these assumptions, Boltzmann's kinetic equation suffices for the description of the
drag resistivity. The final result for the latter quantity 
(and the central result of this paper; see derivation ahead) is
\begin{eqnarray}
\rho_D&=&-\frac{1}{g_0}\frac{\sqrt{\epsilon_{F1}\epsilon_{F2}}}{2^5\pi k_BT}\alpha^2_c
{\cal F}(k_{F2},k_{F1},T,d_c,\alpha_g)\nonumber\\
&\equiv&-\rho_0{\cal F}(k_{F2},k_{F1},T,d_c,\alpha_g)
\label{eq_rho_Drag_final_introd}
\,.
\end{eqnarray}
The reader is referred to the derivation of
Eq. (\ref{eq_rho_Drag_final_nume}) for the definition
of the several quantities in Eq. (\ref{eq_rho_Drag_final_introd}). Using Eq. 
(\ref{eq_rho_Drag_final_introd}) we are able to describe quantitatively  the
drag resistivity measured in graphene at different temperatures \cite{Tutuc2010}. 
Our results are summarized in Fig. \ref{fig_drag_results}.
In that figure the symbols are the experimental data and the dashed lines are the calculated values of $\rho_D$. 
The drag resistivity is represented as function of the gate voltage. The calculated curves have been horizontally
shifted by $V_D=-2$ V, since the graphene in the device had its neutrality point at $V_g=V_D$, whereas the calculated curves
assume neutral graphene when $V_g=0$. The agreement is quantitative up to $V_g=4$; below we discuss the origin of the 
deviations for $V_g<4$ V, which are associated with the electron-puddle formation in the device. To understand the 
experimental conditions  associated with the measurements reproduced in Fig. \ref{fig_drag_results} the reader is referred 
to Ref. \cite{Tutuc2010}. 

The device has two different dielectrics: SiO$_2$ (relative permittivity $\epsilon_b=3.8$),
with thickness $d_b=280$ nm, and Al$_2$O$_3$ (relative permittivity $\epsilon_t=6$ \cite{Tutuc2009}),
with thickness $d_t=14$ nm  (we note that other authors report a value of $\epsilon_{\rm Al_2O_3}=9.1$ \cite{Liao}). 
The latter oxide is sandwiched between the two graphene layers. As detailed in Ref. \cite{Tutuc2010}, 
the densities in the two layers are not independent of each other.
The electronic densities $n_2$  and $n_1$  follow from \cite{Tutuc2010}:
\begin{eqnarray}
\label{eq_electro_1}
 eV_g&=&d_b\frac{e^2}{\epsilon_0\epsilon_b}(n_2+n_1)+v_F\hbar\sqrt{\pi n_2}\,,\\
\label{eq_electro_2}
d_t\frac{e^2}{\epsilon_0\epsilon_t}n_1&=&v_F\hbar(\sqrt{\pi n_2}-\sqrt{\pi n_1})\,.
\end{eqnarray}
The numerical solution of the last two equations gives the electronic
density in the two graphene sheets for any value of $V_g$ (see inset in Fig. \ref{fig_drag_results}).

In our theory, the only free parameter is the intrinsic dielectric of constant graphene or, equivalently,
the interaction parameter: 
\begin{equation}
 \alpha_g=\frac{e^2}{4\pi\epsilon_0v_F\hbar} \approx 2.2\,.
\label{eq_alpha}
\end{equation}
We note that there is some controversy on the value of $\alpha_g$ \cite{Abbamonte}. When graphene is immersed
in a dielectric of relative permittivity $\epsilon_r$, the interaction parameter becomes
$\alpha_r=\alpha_g/\epsilon_r$. In the experimental setup \cite{Tutuc2010} there are three different 
interaction parameters: two associated with 
electron-electron interactions within the two graphene planes
($\alpha_t$ and $\alpha_b$) and one
associated with the inter-plane interaction ($\alpha_c$). 
The constants $\alpha_t$ and $\alpha_b$ play their role in the 
calculation of the dielectric function of the coupled layers.
Explicitly, we have:
\begin{eqnarray}
 \alpha_t&=&2\alpha_g/(1+\epsilon_{\rm Al_2O_3})\,,\\
 \alpha_c&=&\alpha_g/\epsilon_{\rm Al_2O_3}\,,\\
\alpha_b&=&2\alpha_g/(\epsilon_{\rm Al_2O_3}+\epsilon_{\rm SiO_2})\,.
\end{eqnarray}
Taking $\alpha_c$ as reference, we get $\alpha_t\simeq 1.7\alpha_c$
and $\alpha_b\simeq 1.2\alpha_c$; these two numbers are fixed in our model. 
Therefore, the only free parameter is $\alpha_g$, or in alternative
$\alpha_c$. We have chosen $\alpha_c$ as our fitting
parameter and the best fit was obtained for $\alpha_c=0.21$. 
This latter value implies:
\begin{equation}
\alpha_g\simeq 1.3\,, 
\end{equation}
a number close to the one given by Eq. (\ref{eq_alpha}). 

The vertical dashed line  in Fig. \ref{fig_drag_results} sets the limit of validity of our theoretical model in what concerns its application to graphene on dirty substrates.
Indeed, at that point the experimental curves change their curvature from negative to  positive, indicating a 
change of regime. Below $V_g=4$ V,
 the theoretical results start  to deviate from the 
measured data. This happens because the electronic density  in the
bottom layer approaches the regime where  electron-hole 
puddles control the electronic transport in graphene \cite{Tutuc2010}.
In fact, looking at the green curve of Fig. 3.a in Ref. \cite{Tutuc2010}, we see that during the transition
from the regime of electron-electron drag to that of electron-hole drag the behavior of $\rho_D$
is reminiscent of that reported for Hall-measurements in graphene close to the neutrality
point \cite{nov04}. In both cases, the elementary theory, which ignores the effect
of electron-hole puddles, predicts diverging drag resistivity and Hall coefficient. 
We note that our computed curves diverge as the electronic density is reduced;
the divergence is enhanced at high temperatures. In the absence of electron-hole puddles
$\rho_D$ can become large, giving rise to the switching effect we have described in the 
introduction.

\section{Derivation of the drag resistivity formula}

The calculation of the current density $j_2$ (the driven current)
and the electric field $E_1$ (the drag field) requires
the solution of Boltzmann's equation, under the assumption that
the current density in  sheet {\bf 1} is zero, $j_1=0$.
The driven current $j_2$ reads
\begin{equation}
 j_2=E_2g_0\pi v_F\tau_0n_{2}\,.
\label{eq_j2}
\end{equation}
Following a standard procedure \cite{Jauho93,Flensberg94,Flensberg95},
 the drag field $E_1$ is given by:
\begin{equation}
E_1 = \frac{1}{g_0} \frac{e_1g_sg_v} {\pi n_{1} }
\int\frac{d\bm k_1} {4\pi^2}  k_1 \cos\theta_{\bm k_1}
\left.\frac{\partial f_{\bm k_1,c}}{\partial t}\right]_{\rm{ee}}\,,
\label{eq_for_E1_f}
\end{equation} 
where $\left. \partial f_{\bm k_1,c}/\partial t\right]_{\rm{ee}}$
is the collision integral due to inter-plane Coulomb interactions, 
$e_1$ is the charge of the carriers in plane {\bf 1}, $g_0=2e^2/h$,
$n_{1}$ is the electronic density in layer {\bf 1},
and $g_s$ and $g_v$ are the spin and valley degeneracy, respectively.
We stress that unlike previous works \cite{Tse07} we take fully into account
the momentum dependence of the relaxation time. It was noted before 
that ignoring this momentum dependence leads to 
incorrect results \cite{Flensberg95} (for the 2DEG, however, the relaxation time
can be taken momentum independent).
 
The calculation of the collision integral requires the use of the 
Coulomb interaction between electrons in different layers. The latter 
is given by
\begin{eqnarray}
H_{12}&=&\frac{1}{4A}\sum_{\bm k_1,\bm k_2, \bm q}\sum_{\alpha,\alpha',\beta,\beta'}
U_{d_c}(\bm q)f_{\alpha\alpha'}(\bm k_1,\bm q)f_{\beta\beta'}(\bm k_2,\bm q)\times\nonumber\\
&&a^\dag_{\bm k_1 \alpha}a_{\bm k_1+\bm q\alpha'}b^\dag_{\bm k_2\beta}b_{\bm k_2-\bm q\beta'}\,,
\end{eqnarray}
where the creation operators $a^\dag_{\bm k_1 \alpha}$ and $b^\dag_{\bm k_2 \alpha}$
refer to electrons in layers {\bf 1} and {\bf 2}, respectively, $A$ is the 
area of the system, and 
\begin{equation}
 U_{d_c}(\bm q)=\frac{e_1e_2}{2\epsilon_c\epsilon_0q}e^{-q d_c}
\equiv V_c(q)e^{-q d_c}\,,
\end{equation}
where $d_c$ is the inter-layer distance. The transferred  momentum  between the layers is $\bm q=\bm k_1-\bm k_2$,
$\alpha,\alpha',\beta,\beta'=v,c$ refer to the valence ($v$ or -1)
and conduction ($c$ or +1) band electrons, the momentum of electrons in plane
{\bf 1} is denoted by $\bm k_1$ and for electrons in plane {\bf 2} by $\bm k_2$; 
the momentum sums $\bm k_1$ and $\bm k_2$ contain spin summations as well. The chiral
nature of the electron wave function is encoded in the form factors:
\begin{eqnarray}
 f_{\alpha\alpha'}(\bm k_1,\bm q)&=&1 + \alpha\alpha'e^{i(\theta_{\bm k_1}-\theta_{\bm k_1+\bm q})}\,,\\
f_{\beta\beta'}(\bm k_2,\bm q)&=&1 + \beta\beta'e^{i(\theta_{\bm k_2}-\theta_{\bm k_2-\bm q})}\,,
\end{eqnarray}
where $\theta=\arctan(k_y/k_x)$.

Following a standard approach \cite{ziman,Jauho93,Flensberg94,Flensberg95}, 
we have 
\begin{eqnarray}
\int\frac{d\bm k_1}{4\pi^2} k_1\cos\theta_{\bm k_1}
\left.\frac{\partial f_{\bm k_1,c}}{\partial t}\right]_{\rm{ee}}=
-E_2\frac{\pi e v_F\tau_0}{2\hbar k_BT} \times\nonumber\\
\int\frac{d\bm q}{4\pi^2}\int \hbar d\omega
\frac{\vert U(\bm q,\omega)\vert^2}{\sinh^2(\hbar\omega/2k_BT)}
P_{1c}(\bm q,\omega)P_{2c}(\bm q,\omega)\,,
\end{eqnarray}
where $U(\bm q,\omega) = U_{d_c}(q)/\epsilon(\bm q,\omega) $
 is the dynamic Coulomb interaction,
evaluated using the random phase approximation  for the dynamic dielectric function 
$\epsilon(\bm q,\omega)$.  
We note that $\tau_0$ is not the relaxation time (see above), and it 
 cancels out when the ratio $E_1/j_2$ is computed.
We also have
\begin{eqnarray}
\label{eq_PJ_def}
 P_{jc}(q,\omega)=\int\frac{d\bm k_1}{4\pi^2}G^{(j)}_{\bm k_j,\bm q}
\delta f_{\bm k_j,\bm q}\delta(\hbar\omega+\hbar\omega_{\bm k_j,\bm q})\,,
\end{eqnarray}
where 
$\delta f_{\bm k_j,\bm q}=f_{\bm k_j+\bm q/2}-f_{\bm k_j-\bm q/2}$,
$\hbar\omega_{\bm k_j,\bm q}=
\epsilon_{\bm k_j-\bm q/2}-\epsilon_{\bm k_j+\bm q/2}$ ($j=1,2$),
and
the function $G^{(j)}_{\bm k_1,\bm q}$, with $j=1,2$, is defined as
\begin{eqnarray}
G^{(j)}_{\bm k_j,\bm q}&=&\frac{1}{2+2\delta_{1,j}}\left[
\vert\bm k_j-\bm q/2\vert\cos\theta_{\bm k_j-\bm q/2}
\right.
\nonumber\\
&-&
\left.
\vert\bm k_j+\bm q/2\vert\cos\theta_{\bm k_j+\bm q/2}]
\right]
\vert f_{cc}(\bm k_j,\bm q)\vert^2
\,.
\label{eq_Gfunctions}
\end{eqnarray}
The difference between our calculation and that of Tse {\it et al.} \cite{Tse07}
is precisely in the form of the $G^{(j)}_{\bm k_j,\bm q}$. The dynamic dielectric function 
$\epsilon(\bm q,\omega)$ for the two-layer system is given by \cite{Flensberg95,Uchoa}:
\begin{equation}
\epsilon(\bm q,\omega)=\epsilon_2(\bm q,\omega)\epsilon_1(\bm q,\omega)-
U^2_{d_c}(q){\cal P}_1(\bm q,\omega){\cal P}_2(\bm q,\omega)\,,  
\end{equation}
where $\epsilon_j(\bm q,\omega)=1-V_j(\bm q){\cal P}_j(\bm q,\omega)$ 
and ${\cal P}_j(\bm q,\omega)$
are the dielectric and polarization functions
of each of the individual layers \cite{Wunsch}. 
Notice that $\alpha_b$ and $\alpha_t$ enter in
$\epsilon_b(\bm q,\omega)$ and $\epsilon_t(\bm q,\omega)$,
respectively. Once the functions
$ P_{jc}(q,\omega)$ are computed, a relatively simple expression
follows for the drag resistivity:
\begin{eqnarray}
\rho_D=-\frac{1}{g_0}\frac{\sqrt{\epsilon_{F1}\epsilon_{F2}}}{2^5\pi k_BT}\alpha^2_c
{\cal F}(k_{F2},k_{F1},T,d_c,\alpha_g)
\,,
\label{eq_rho_Drag_final_nume}
\end{eqnarray}
which leads to Eq. (\ref{eq_rho_Drag_final_introd}) and
$\alpha_c$ is the interaction parameter
of graphene considering the dielectric in between the two graphene layers
made out of Al$_2$O$_3$. 
\begin{figure}[ht]
 \begin{center}
\includegraphics*[width=6cm]{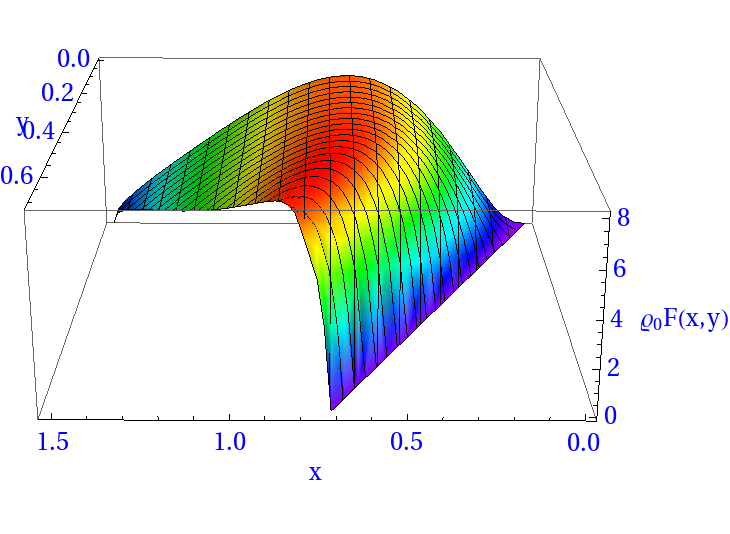}
\includegraphics*[width=6cm]{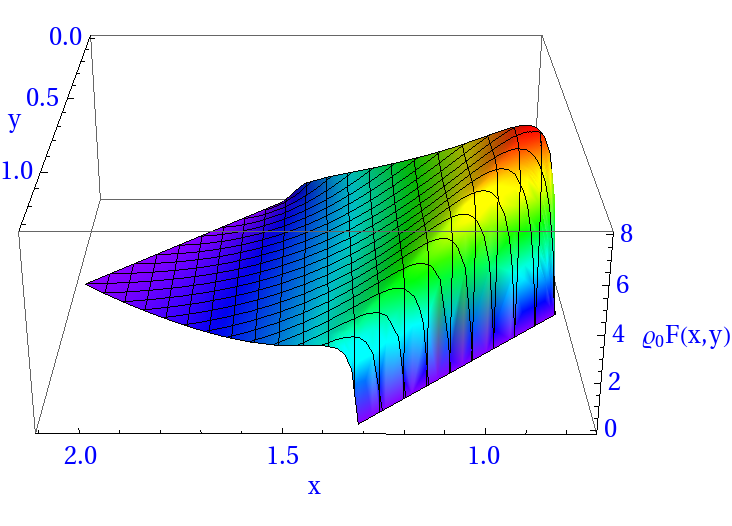}
\includegraphics*[width=6cm]{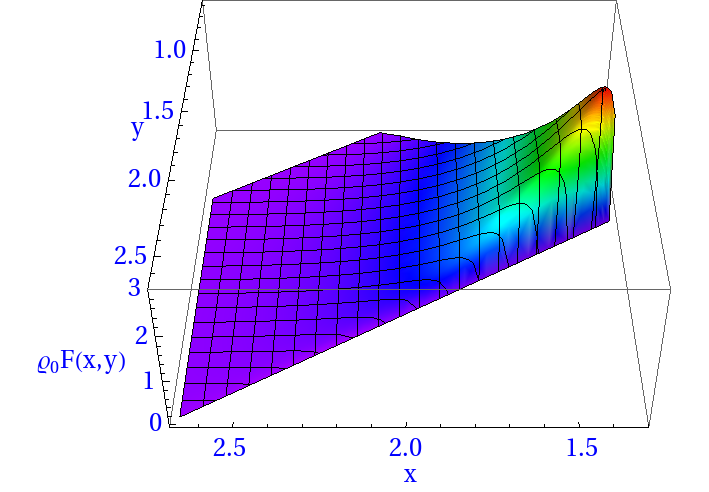}
 \end{center}
\caption{ Plot of the integration kernel $\rho_0F(x,y)$ in 
Eq. (\ref{eq_Rho_Drag_final}) for the parameters
$V_{g}=3$ V,  $d=14$ nm, $\alpha_g=0.21$, and $T=300$ K. The three figures 
correspond to three different regions of the $(q,\Omega)$ plane.
From top to bottom the regions are ordered as $m=$1, $m=$2, and $m=$3 (see text).}
\label{fig_drag_Kernel}
\end{figure}
The dimensionless function ${\cal F}(k_{F2},k_{F1},T,d_c,\alpha_g)$ is defined as:
\begin{equation}
 {\cal F}(k_{F2},k_{F1},T,d_c,\alpha_g)=\sum_{m=1}^3
\int_0^{x_{\rm m}}dx 
\int_0^{y_{\rm m}} dy
F(x,y)
\,,
\label{eq_Rho_Drag_final}
\end{equation}
where
\begin{equation}
F(x,y)=\frac{x^9(x^2-y^2)}
{\vert\epsilon(x,y)\vert^2}
\frac{\Phi_1(x,y)\Phi_2 (x,y)} 
{\sinh^{2}\frac{v_F\hbar\sqrt{k_{F1}k_{F2}}y}{2k_BT}}
e^{-2d_cx\sqrt{k_{F1}k_{F2}}}
\,,
\end{equation}
and $x$ and $y$ are dimensionless variables defined as 
$q=\sqrt{k_{F1}k_{F2}}x$ and $\omega=v_F\sqrt{k_{F1}k_{F2}}y$, respectively.
The three integration regions are: $0<x_1<b$, $0<y_1<{\rm min}(x,-x+b)$; 
$b/2<x_2<(a+b)/2$, ${\rm max}(-x+b,x-b)<y_2<{\rm min}(x,-x+a)$; 
$a/2<x_3<\infty$, ${\rm max}(-x+a,x-b)<y_3<q$
where $a=2\sqrt{k_{F1}/k_{F2}}$ and $b=2\sqrt{k_{F2}/k_{F1}}$.
The function $\epsilon(x,y)$ is defined as:
\begin{eqnarray}
 \epsilon(x,y)&=&
(x\sqrt{x^2-y^2}+P_{12})(x\sqrt{x^2-y^2}+P_{21})
\nonumber\\
&-&
P_{12}P_{21}e^{-2d_c\sqrt{k_{F1}k_{F2}}x}
\,,
\end{eqnarray}
where
\begin{eqnarray}
P_{sj}=4\alpha_s\sqrt{\frac{k_{Fs}}{k_{Fj}}}\sqrt{x^2-y^2}
-i\frac{\alpha_s}{2}x^2\Phi_s(x,y)
\,,
\end{eqnarray}
and $\alpha_s$ equals $\alpha_b$ or $\alpha_t$, depending on the graphene layer.
Below we outline the calculation of the functions $\Phi_s(x,y)$.

In Fig. \ref{fig_drag_Kernel} we plot the quantity $\rho_0F(x,y)$ for particular
values of $k_{Fi}$ (as determined by the gate voltage), the effective 
interaction parameter $\alpha_c$, the distance between the two
layers, and the temperature. In that figure the three panels, from top to bottom,
represent the contribution of the three integration domains 
defined by $x_m$ and $y_m$. At low temperatures and high back-gate voltage,
only the contribution of region $m=1$ is significant; on the other hand,
at high temperature and low densities all 
the three regions ($m=1,2,3$) give a significant contribution to $\rho_D$, as shown 
with the example in Fig. \ref{fig_drag_Kernel}.

\section{Particular limits of $\rho_D$}

\begin{figure}[ht]
 \begin{center}
\includegraphics*[width=8cm]{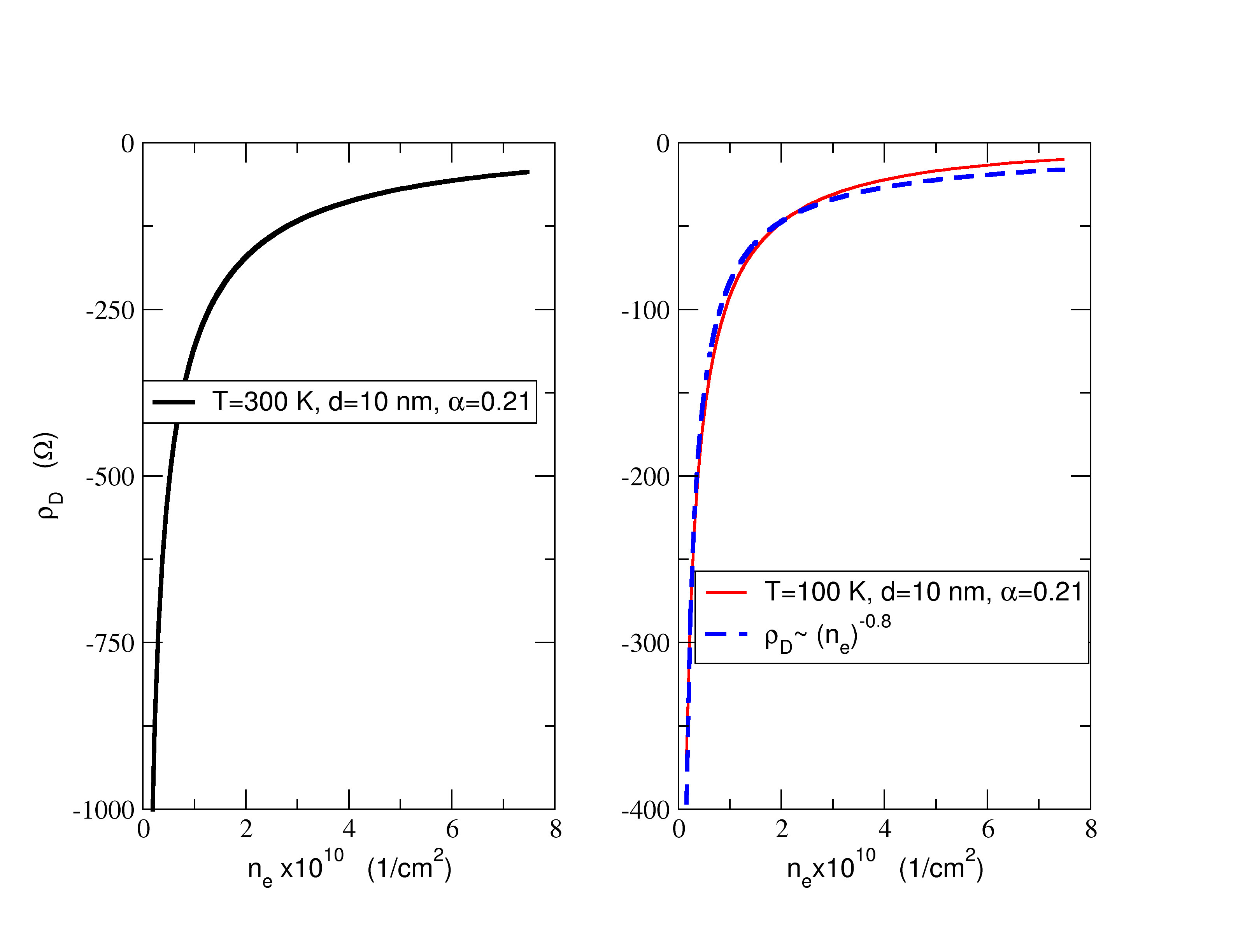}
 \end{center}
\caption{Dependence of the drag resistivity on the density
for two different temperatures, $T=300$ K and $T=100$ K.
The other parameters are $\alpha_b=\alpha_c=\alpha_t=0.21$,
$d_b=d_t=280$ nm, and $d_c=10$ nm. The dashed curve is a fit 
to the expression $\rho_D\propto n_e^{-0.8}$.}
\label{fig_Rho_D_versus_n_and_d}
\end{figure}

Let us now consider the case where 
$\alpha_b$, $\alpha_c$, and $\alpha_t$ are all equal, as 
is the electronic density in the two sheets. Taking the limit 
of very low densities, we see that $\rho_D$ can reach values of the 
order of $10^3$ $\Omega$, as in the case represented in Fig. 
\ref{fig_Rho_D_versus_n_and_d}.
It is the possibility of having large values of $\rho_D$
that suggests the mechanism for turning {\it on}-and-{\it off}
the electric current in a double-layer graphene-based transistor.
We also note that in the density range $n_e\sim 3\times10^{12}$ cm$^{-2}$
to  $n_e\sim 0.5\times10^{12}$ cm$^{-2}$,
$\rho_D$ is better described by $\rho_D\propto n_e^{-\alpha(n_e)}$,
where the exponent $\alpha(n_e)$ shows a crossover from $\alpha(n_e)\simeq2$
to $\alpha(n_e)\simeq1$ as the electronic density is reduced.

Finally, when $\epsilon_F\gg k_BT$ and $d_ck_F\gg 1$,
 the asymptotic regime for $\rho_D$ reads
\begin{equation}
 \rho_D\propto -\frac{1}{g_0}\frac{(k_BT)^2}{v_F^2\hbar^2n_e^4d_c^6}\,.
\label{eq_high_density}
\end{equation}
The asymptotic formula (\ref{eq_high_density})
has a different dependence on $d$ and $n_e$ 
from that found in  Ref. \cite{Tse07}, due to the 
inclusion of the momentum dependence in the scattering time; in Ref. \cite{Tse07}
the integration over $q$ has a kernel of the form $q^3/\sinh^2(qd_c)$, whereas 
our kernel, in the same limit, has the form 
$q^5/\sinh^2(qd_c)$. Thus we obtain $\rho_D\propto d^6$ and Tse {\it et al.}
have obtained $\rho_D\propto d^4$. The difference by a factor of two in the power 
of the momentum $q$ comes from: (i) $\tau_{\bm q}\propto q$; (ii)
$\tau_{\bm q}$ appears twice in the right-hand side of Eq. (\ref{eq_for_E1_f}).
It is worth stressing that
Eq. (\ref{eq_high_density})
 does not apply to the conditions of the experiment
we have described in Fig. \ref{fig_drag_results}, and 
the experimental conditions where it would apply can hardly be reached.
Also, the values of $\rho_D$ of an experiment done in the 
regime of validity of Eq. (\ref{eq_high_density}) 
would be difficult to measure.

\section{Technical details: Calculation of the functions $\Phi_j(x,y)$ at zero temperature}

The functions $\Phi_j(x,y)$ -- $j=1,2$ -- depend on the integration limits $x_i$ and 
$y_i$ -- $i=1,2,3$; we also define $\Omega=\omega/v_F$. We write 
$\Phi_j(q,\Omega,k_{Fj})=\Phi_j^+(q,\Omega,k_{Fj})+\Phi_j^-(q,\Omega,k_{Fj})$, where
\begin{equation}
 \Phi_j^\pm(q,\Omega,k_{Fj})=
\int_{-\infty}^\infty 
d\phi\sinh^2\phi f_j[(q\cosh^2\phi\pm\Omega)/2]\,.
\label{eq_phi_def}
\end{equation}
At zero temperature we have:
\begin{equation}
f_j[(q\cosh^2\phi\pm\Omega)/2] = \theta[2k_{Fj}-(q\cosh^2\phi\pm\Omega)]\,. 
\end{equation}
In terms of the functions  $\Phi_j^+(q,\Omega,k_{Fj})$ and $\Phi_j^-(q,\Omega,k_{Fj})$,
 the function $\Phi_j(q,\Omega,k_{Fj})$ reads: 
\begin{enumerate}
\item {\bf region 1:} $0<q<k_{Fj}$ and $0<\Omega<q$, in which case we have
$\Phi_j=\Phi_j^+-\Phi_j^-$ (we have omitted the arguments of the functions
$\Phi_j(q,\Omega,k_{Fj})$
for the sake of simplicity);
\item {\bf region 2:} $k_{Fj}<q<2k_{Fj}$ and $0<\Omega<2k_{Fj}-q$, in which case we have
$\Phi_j=\Phi_j^+-\Phi_j^-$;
\item {\bf region 3:} $k_{Fj}<q<2k_{Fj}$ and $2k_{Fj}-q<\Omega<q$, in which case we have
$\Phi_j=-\Phi_j^-$;
\item {\bf region 4:} $q>2k_{Fj}$ and $q-2k_{Fj}<\Omega<q$, in which case we have
$\Phi_j=-\Phi_j^-$.
\end{enumerate}
In the regions where $\Phi_j^\pm$ is finite we have
\begin{eqnarray}
 \Phi_j^+(q,\Omega,k_{Fj})&=&-\phi_{\rm min} +\frac{1}{2} \sinh(2\phi_{\rm min})\,,\\
 \Phi_j^-(q,\Omega,k_{Fj})&=&-\phi_{\rm max} +\frac{1}{2} \sinh(2\phi_{\rm max})\,,
\end{eqnarray}
where $\phi_{\rm max}=\acosh[(2k_{Fj}+\Omega)/q]$ and $\phi_{\rm min}=\acosh[(2k_{Fj}-\Omega)/q]$.
\begin{figure}[ht]
 \begin{center}
   \includegraphics[width=8cm]{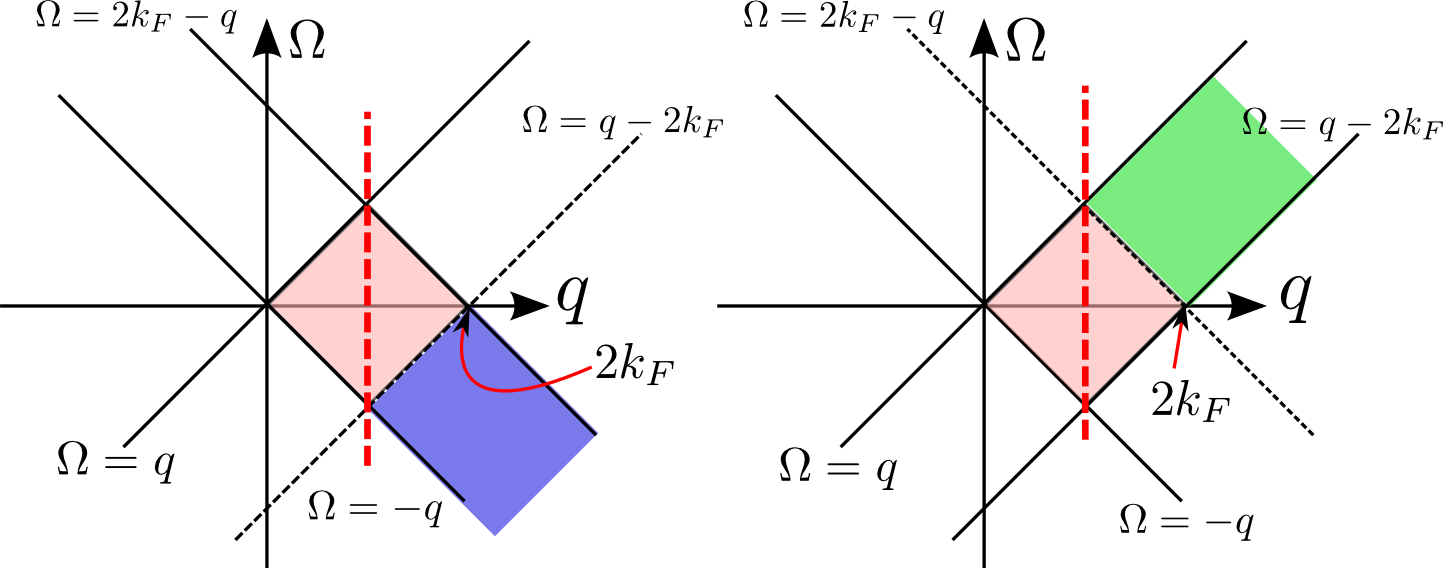}
 \end{center}
\caption{Domains where the functions 
$\Phi_j^+(q,\Omega,k_{Fj})$ (left) 
and 
$\Phi_j^-(q,\Omega,k_{Fj})$ (right) are finite. In the triangular area,
both functions are finite. In the rectangular area only one of the functions is
finite.}
\label{fig_domains}
\end{figure}
The four regions above have their graphical representation in Fig. \ref{fig_domains}.
In the triangular area of the diagram \ref{fig_domains}, the polarization
reads
\begin{equation}
{\cal P}_j(\bm q,\Omega)=-\frac{2k_{Fj}}{\pi v_F\hbar }+\frac{i}{4\pi v_F\hbar}\frac{q^2}{\sqrt{q^2-\Omega^2}}\Phi_j(\bm q,\Omega)\,. 
\label{eq_polariz_triang}
\end{equation}
In the rectangular area of the same diagram, the polarization is obtained by adding 
 to Eq. (\ref{eq_polariz_triang}) the term
\begin{equation}
 \frac{1}{4\pi v_F\hbar}\frac{1}{\sqrt{q^2-\Omega^2}}\Theta_j(\bm q,\Omega)\,,
\end{equation}
where $\Theta_j(\bm q,\Omega)=\Theta_a(\bm q,\Omega)+\Theta_b(\bm q,\Omega)$
and 
\begin{eqnarray}
 \Theta_a(\bm q,\Omega) &=& (2k_{Fj}-\Omega)\sqrt{q-2k_{Fj}+\Omega}\times\nonumber\\
&&\sqrt{q+2k_{Fj}-\Omega}\,,\\
 \Theta_b(\bm q,\Omega) &=& -2q^2\arctan\sqrt{\frac{q+\Omega-2k_{Fj}}{q-\Omega+2k_{Fj}}}\,.
\end{eqnarray}
We have used these results for the polarization in the calculation of the dynamic
dielectric function.

\section{Conclusions}
We have given a quantitative theory of Coulomb drag in graphene. We have shown that for ultra-clean
graphene, unlike the case of a conventional 2DEG, can be tuned to a region of diverging drag and
longitudinal resistivity allowing the use of double layer structures for device applications
where the on-and-off ratio for the current flow has to be large. We also show that our theory
explains quantitatively the experimental data for the drag resistivity in dirty devices away
from the electron-hole puddle region. As this region shrinks in cleaner devices, the validity of
our theory extends. 

\section{Supplementary Information}
\begin{figure}[ht]
 \begin{center}
   \includegraphics[width=8cm]{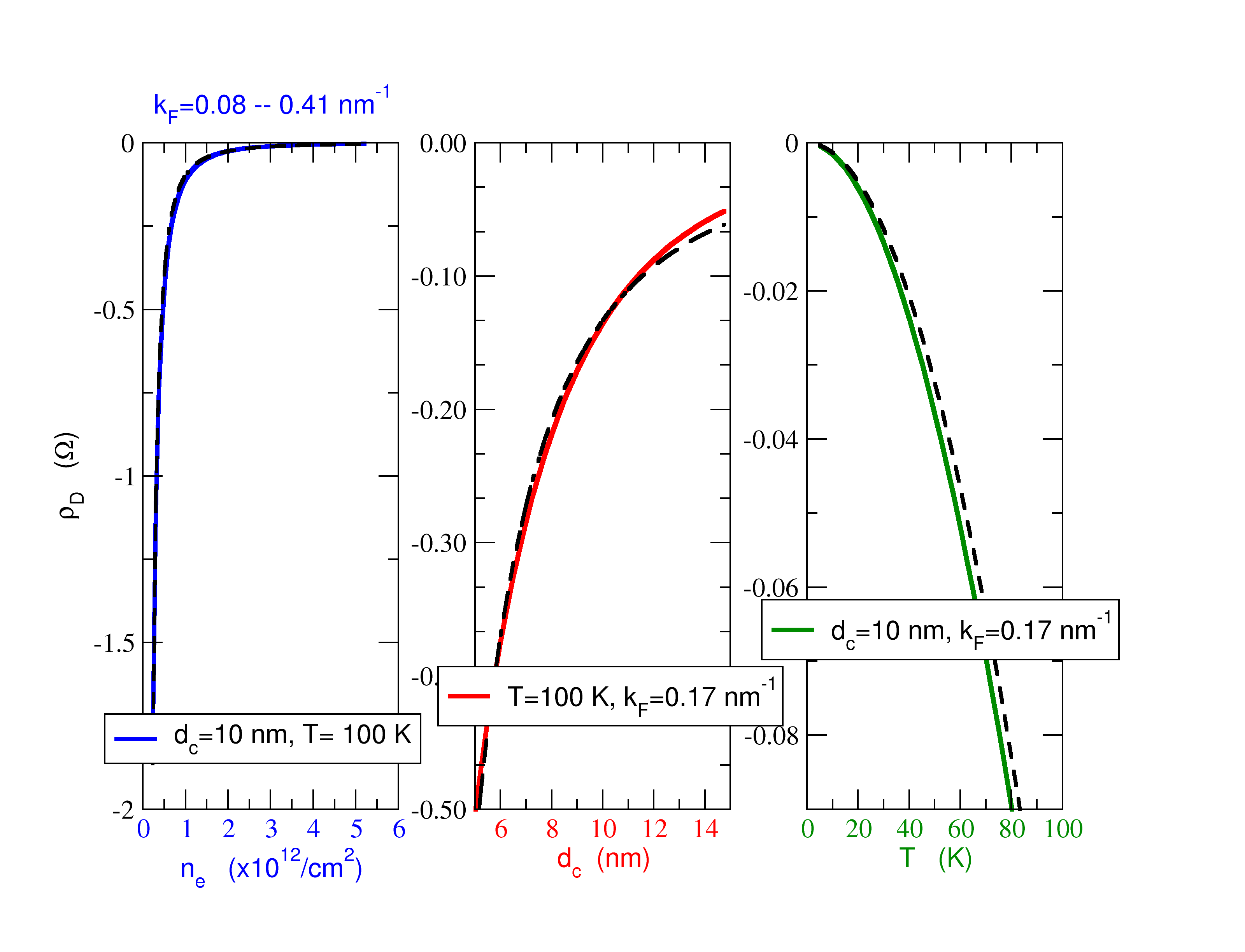}
 \end{center}
\caption{Dependence of the drag resistivity as function of the electronic density, $n_e$,
the interlayer distance, $d_c$, and the temperature, $T$. The used value 
of $\alpha$ is $\alpha=0.21$}
\label{fig_domains_2nd}
\end{figure}

In Fig. \ref{fig_domains_2nd} we plot the dependence of the drag resistivity 
as function of   the electronic density, $n_e$,
the interlayer distance, $d_c$, and the temperature, $T$. The density range
scanned in this figure is different from that given in Fig. \ref{fig_Rho_D_versus_n_and_d}.
In the three panels of Fig. \ref{fig_domains_2nd} the black dashed lines
are fits to a power law of the form $\rho_D\sim x^{\alpha_x}$,
where $x$ represents one of three parameters $n_e$, $d_c$, and $T$. From these
fits we have found that $\rho_D$ follows, in the regime of parameters of Fig.
\ref{fig_domains_2nd}, roughly the behavior
\begin{equation}
 \rho_D\sim \frac{(Tk_B)^2}{v_F^2\hbar^2 n_e^2 d_c^2}\,.
\label{eq_rhoD_about}
\end{equation}
It is also clear from Fig. \ref{fig_domains_2nd} that Eq. (\ref{eq_rhoD_about})
holds only approximately. Also, comparing the dependence of $\rho_D$ on $n_e$, as given
in the regime of parameters of Fig. \ref{fig_Rho_D_versus_n_and_d},
with that of Eq. (\ref{eq_rhoD_about}), obtained from Fig. \ref{fig_domains_2nd},
we see that the exponent $\alpha_{n_e}$ changes with density, being smaller at
smaller density values.

%

\section{Acknowledgments}
We thank useful discussions with A. Geim and K. Novoselov. 
AHCN acknowledges DOE grant DE-FG02-08ER46512 and ONR grant MURI
N00014-09-1-1063.

\bibliographystyle{eplbib} 

\end{document}